\PassOptionsToPackage{sort&compress}{natbib}

\documentclass[12pt,twoside]{article}   
\usepackage[super,sort,comma]{natbib}
\usepackage{comment}
\usepackage{physics}
\usepackage{mathtools}
\usepackage{fancyhdr}		

\usepackage[section]{placeins}   %

\usepackage{graphicx}

\usepackage{caption}
\usepackage{subcaption}

\usepackage{float}

\makeatletter \renewcommand\@biblabel[1]{$^{#1}$} \makeatother
\newcommand{\cen}[1]{\begin{center} #1 \end{center}}

\usepackage[mathlines]{lineno}

\usepackage{hyperref}
\hypersetup{ colorlinks,
    citecolor=blue,
    filecolor=blue,
    linkcolor=blue,
    urlcolor=blue
}

\usepackage{xcolor}

\definecolor{gray}{rgb}{0.6,0.6,0.6}
\definecolor{red}{rgb}{0.85,0,0}
\definecolor{green}{rgb}{0.0, 0.42, 0.24}
\definecolor{blue}{rgb}{0,0,0.85}
\definecolor{beige}{rgb}{0.92,0.87,0.78}
\definecolor{darkviolet}{rgb}{0.58, 0.0, 0.83}

\usepackage{soul}
\usepackage[all]{hypcap}    

\usepackage[separate-uncertainty=true,
            bracket-numbers=false,
            range-units=single,
            list-units=single,
            per-mode=symbol]
            {siunitx}
\DeclareSIUnit{\nkeV}{\keV\tothe{-1}}
\DeclareSIUnit{\MeVu}{\MeV\per\atomicmassunit}
\DeclareSIUnit{\nm}{\nano\meter}
\DeclareSIUnit{\um}{\micro\meter}
\DeclareSIUnit{\squm}{\micro\meter\squared}
\DeclareSIUnit{\cbum}{\cubic\micro\meter}
\DeclareSIUnit{\ncbum}{\micro\meter\tothe{-3}}
\DeclareSIUnit{\nsqum}{\micro\meter\tothe{-2}}
\DeclareSIUnit{\keVum}{\keV\per\um}
\DeclareSIUnit{\Gy}{\gray}
\DeclareSIUnit{\nGy}{\per\gray}
\DeclareSIUnit{\sqGy}{\square\gray}
\DeclareSIUnit{\kg}{\kilogram}

\usepackage{amsmath}
\usepackage{amsthm}
\usepackage{amssymb}
\usepackage[noabbrev]{cleveref}

\usepackage[version=4]{mhchem}
\usepackage{parskip}

\newcommand{\Xk}[3]{{}^{#2}{#1}_{\mathrm{#3}}}

\newcommand{\aG}{\alpha_\mathrm{G}}
\newcommand{\bG}{\beta_\mathrm{G}}

\newcommand{\ci}{\mathrm{c}_i}
\newcommand{\ck}{\mathrm{c}_k}
\newcommand{\cK}{\mathrm{c}_K}
\newcommand{\cN}{\mathrm{c}_N}
\newcommand{\ti}{\mathrm{t}_i}
\newcommand{\tN}{\mathrm{t}_N}
\newcommand{\tk}{\mathrm{t}_k}

\newcommand{\Zchem}{\tilde{Z}}

\newcommand{\nE}{n^*}

\newcommand{\Gr}{G_\mathrm{r}}

\newcommand{\ie}{i.e.,~}
\newcommand{\eg}{e.g.,~}

\usepackage{booktabs}

\usepackage{siunitx}

\usepackage[font=normalsize,labelfont={bf}]{caption}
\newtheoremstyle{hypoth}
  {5pt}
  {5pt}
  {}
  {}
  {\bfseries}
  {}
  {\newline}
  {}

\theoremstyle{hypoth}
\newtheorem{hypoth}{Hypothesis}

\newtheoremstyle{simp}
  {5pt}
  {5pt}
  {}
  {}
  {\bfseries}
  {}
  {\newline}
  {}

\theoremstyle{simp}
\newtheorem{simp}{Simplification}

\newcommand{\simplification}[3]{%
\begin{simp}[\bf#1]\label{#2}%
 #3%
\end{simp}
}

\newtheoremstyle{approxim}
  {5pt}
  {5pt}
  {}
  {}
  {\bfseries}
  {}
  {\newline}
  {}

\theoremstyle{approxim}
\newtheorem{approxim}{Approximation}

\newcommand{\approximation}[3]{%
\begin{approxim}[\bf#1]\label{#2}%
 #3%
\end{approxim}
}

\newtheoremstyle{cond}
  {5pt}
  {5pt}
  {}
  {}
  {\bfseries}
  {}
  {\newline}
  {}

\theoremstyle{cond}
\newtheorem{cond}{Condition}

\newcommand{\condition}[3]{%
\begin{cond}[\bf#1]\label{#2}%
 #3%
\end{cond}
}

\newbox\savedeqs 
\makeatletter 
\newcommand\saveandprinteq[1]{
  \begingroup
    \expandafter\let\csname \@currenvir\expandafter\endcsname\csname listeq@\@currenvir\endcsname
    \expandafter\let\csname end\@currenvir\expandafter\endcsname\csname listeq@end\@currenvir\endcsname
    \edef\listeq@temp{
      \noexpand\begin{\@currenvir}%
        \unexpanded{#1}%
      \noexpand\end{\@currenvir}%
    }%
    \savecounters@ 
      \global\setbox\savedeqs=\vbox{\unvbox\savedeqs\listeq@temp}
    \restorecounters@ 
    \listeq@temp 
  \endgroup
}
\newcommand*\listeqpatch[1]{
  \expandafter\let\csname listeq@#1\expandafter\endcsname\csname #1\endcsname
  \expandafter\let\csname listeq@end#1\expandafter\endcsname\csname end#1\endcsname
  \renewenvironment{#1}{\collect@body\saveandprinteq}{}%
}

\makeatother 

\listeqpatch{equation}
\listeqpatch{align}
\listeqpatch{gather}
\listeqpatch{multline}
\listeqpatch{flalign}

\usepackage{draftwatermark}

\SetWatermarkText{Preprint version}
\SetWatermarkColor[gray]{0.5}
\SetWatermarkFontSize{1cm}
\SetWatermarkAngle{90}
\SetWatermarkHorCenter{20cm}

\begin{document}

\thispagestyle{empty}
\begin{center}
\textbf{\Large{Biophysical modeling of low-energy ion irradiations with NanOx}}
\end{center}
\topskip0pt
\vspace*{\fill}
\large{This is the pre-peer reviewed version of the following article: \textit{Alcocer-Ávila M, Levrague V, Delorme R, Testa E, and Beuve M. Biophysical modeling of low-energy ion irradiations with NanOx. Med Phys. 2024; 1-14}, which has been published in final form at \href{https://doi.org/10.1002/mp.17407}{https://doi.org/10.1002/mp.17407}. This article may be used for non-commercial purposes in accordance with Wiley Terms and Conditions for Use of Self-Archived Versions.}
\vspace*{\fill}
\newpage

\cen{\sf {\Large {\bfseries Biophysical modeling of low-energy ion irradiations with NanOx} \\  
\vspace*{10mm}
Mario Alcocer-Ávila\,$^{1, \dagger}$, Victor Levrague\,$^{2,\dagger}$, Rachel Delorme\,$^{2}$, Étienne Testa\,$^{1}$ and Michaël Beuve\,$^{1,*}$} \\
\vspace*{5mm}
$^{1}$Université Claude Bernard Lyon 1, CNRS/IN2P3, IP2I Lyon, UMR 5822, Villeurbanne, F-69100, France\\
$^{2}$University of Grenoble Alpes, CNRS, Grenoble INP, LPSC-IN2P3, 38000 Grenoble, France
\vspace{5mm}\\
}

\pagenumbering{roman}
\setcounter{page}{1}
\pagestyle{plain}
$\dagger$ These authors share first authorship.\\
$^*$Author to whom correspondence should be addressed: michael.beuve@univ-lyon1.fr \\

\begin{abstract}
\noindent
{{\bf Background:} Targeted radiotherapies with low-energy ions show interesting possibilities for the selective irradiation of tumor cells, a strategy particularly appropriate for the treatment of disseminated cancer. Two promising examples are boron neutron capture therapy (BNCT) and targeted radionuclide therapy with $\alpha$-particle emitters (TAT). The successful clinical translation of these radiotherapies requires the implementation of accurate radiation dosimetry approaches able to take into account the impact on treatments of the biological effectiveness of ions and the heterogeneity in the therapeutic agent distribution inside the tumor cells. To this end, biophysical models can be applied to translate the interactions of radiations with matter into biological endpoints, such as cell survival.} \\ 
{{\bf Purpose:} The NanOx model was initially developed for predicting the cell survival fractions resulting from irradiations with the high-energy ion beams encountered in hadrontherapy. We present in this work a new implementation of the model that extends its application to irradiations with low-energy ions, as the ones found in TAT and BNCT.} \\
{{\bf Methods:} The NanOx model was adapted to consider the energy loss of primary ions within the sensitive volume (i.e. the cell nucleus). Additional assumptions were introduced to simplify the practical implementation of the model and reduce computation time. In particular, for low-energy ions the narrow-track approximation allowed to neglect the energy deposited by secondary electrons outside the sensitive volume, increasing significantly the performance of simulations. Calculations were performed to compare the original hadrontherapy implementation of the NanOx model with the present one in terms of the inactivation cross sections of human salivary gland (HSG) cells as a function of the kinetic energy of incident $\alpha$-particles.} \\
{{\bf Results:} The predictions of the previous and current versions of NanOx agreed for incident energies higher than 1 MeV/n. For lower energies, the new NanOx implementation predicted a decrease in the inactivation cross sections that depended on the length of the sensitive volume.} \\
{{\bf Conclusions:} We reported in this work an extension of the NanOx biophysical model to consider irradiations with low-energy ions, such as the ones found in TAT and BNCT. The excellent agreement observed at intermediate and high energies between the original hadrontherapy implementation and the present one showed that NanOx offers a consistent, self-integrated framework for describing the biological effects induced by ion irradiations. Future work will focus on the application of the latest version of NanOx to cases closer to the clinical setting.} \\

\subsubsection*{KEYWORDS} NanOx, biophysical model, cell survival probability, targeted radionuclide therapy, boron neutron capture therapy, \mbox{$\alpha$-particles}, low-energy ions
\end{abstract}

\newpage

\tableofcontents

\newpage

\setlength{\baselineskip}{0.7cm}      

\pagenumbering{arabic}
\setcounter{page}{1}
\pagestyle{fancy}


\section{Introduction}
\label{sec:intro}
The number of new cancer patients will continue to increase worldwide in the near future \cite{sung2021}. It is estimated that more than half of cancer patients will undergo radiotherapy at some point \cite{thiagarajan2021,borras2016}. Many recent developments in radiotherapy have focused on exploiting the interesting energy deposition features of high-energy ions in the context of hadrontherapy, mainly through the use of proton and carbon ion beams to achieve a more conformal dose to tumors, a better normal tissue sparing and an increased relative biological effectiveness (RBE) compared to photon beams. Hadrontherapy  is considered particularly advantageous for the treatment of deep-seated and radioresistant local tumors, though further investigation is still needed to confirm that its dosimetric advantages translate into clinical benefits over conventional X-ray radiotherapy \cite{rossi2022,balosso2022}. 

On the other hand, innovative radiotherapy techniques involving the internal targeted irradiation of tumors with low-energy ions have reemerged in recent years. The two best-known examples are boron neutron capture therapy (BNCT) and targeted radionuclide therapy with $\alpha$-emitters (also known as targeted alpha therapy, TAT). In contrast to hadrontherapy, the systemic and targeted nature of these approaches makes them attractive for treating patients with metastatic disease.
 
In BNCT, a non-radioactive boron \big(\ce{^{10}B}\big) compound having a high specificity for tumor cells is injected into the patient. Then, the tumor is irradiated with an external source of epithermal neutrons. The thermalized neutrons interact with the boron nuclei, triggering the \ce{^{10}B}(n,$\alpha$)\ce{^{7}Li} neutron capture reaction, with a high cross section\cite{prosdocimi1963precise}. 
The latter reaction emits 1.77 MeV $\alpha$-particles and 1.02 MeV \ce{^{7}Li} ions at a branching ratio of 6.3\%; and 1.47 MeV $\alpha$-particles, 0.84 MeV \ce{^{7}Li} ions and 0.478 MeV photons at a branching ratio of 93.7\% \cite{sato2018}. These ions have very high linear energy transfer (LET), between 175 and 190 keV/µm \cite{astar2005} or between 357 and 380 keV/µm \cite{ziegler2003} for $\alpha$-particles or \ce{^{7}Li} ions, respectively. This characteristic is accompanied by very short ranges in water of less than 9 µm\cite{vanvliet-vroegindeweij2001a,ziegler2003,fukunaga2020,suzuki2020}, which limits damage within the diameter of a single tumor cell, largely sparing surrounding healthy tissues \cite{seneviratne2022}. In addition, other nuclear reactions may also occur with the nuclei of elements commonly present in tissues, e.g., hydrogen and nitrogen: \ce{^{1}H}(n,n)p, \ce{^{1}H}(n,$\gamma$)\ce{^{2}H} and \ce{^{14}N}(n,p)\ce{^{14}C}. These reactions will contribute to the total absorbed dose through the generated photons and protons and must be considered in treatment planning \cite{jin2022,sato2018}. Some of the factors leading to a renewed interest in BNCT in the last decades are the development of boron compounds with high specificity, accelerator-based neutron sources, and treatment planning software \cite{lamba2021,jin2022,suzuki2020,cartelli2020}. To date, BNCT has been investigated in phase I/II clinical trials in a variety of disease sites, including glioblastoma multiforme, head and neck cancers, hepatocellular carcinoma and melanomas \cite{malouff2021}. Altogether, these advances make the transition from experimental reactor-based facilities to hospitals more feasible, reinforcing the need to provide tools capable of predicting the BNCT treatment efficacy including the great complexity of the dose contributions even at microscale.

Besides, TAT is a cancer treatment modality in which a radiopharmaceutical, consisting of an $\alpha$-emitter coupled to a suitable carrier molecule by means of bifunctional chelating agents, is injected into the patient. The carrier molecule (\eg a monoclonal antibody, a peptide or an oligonucleotide) is able to target the tumor cells or the tumor microenvironment to closely irradiate the tumor, effectively killing malignant cells while minimizing the absorbed dose to healthy tissues. The $\alpha$-particles emitted in radioactive decays have energies ranging from about 4 to 9 MeV, short path length ($<$ \SI{100}{\um}, equivalent to a few cell diameters) and high LET (60--230 \si{\keVum}). The latter makes $\alpha$-emitters very effective in inducing complex DNA damage, leading to a higher cytotoxicity than with $\gamma$ and $\beta$-emitters, irrespective of the presence of oxygen. Moreover, while the short path length of $\alpha$-particles is considered particularly advantageous for the treatment of systemic disease and micrometastases, their therapeutic benefit may be extended to macroscopic lesions via off-target effects (\ie bystander effect and immune responses)\cite{guerraliberal2020,pouget2021}. As with BNCT, the concept of TAT has been investigated for many years. To date, however, only one $\alpha$-emitter, radium-223 \big(\ce{^{223}Ra}\big) has been approved for clinical use. A few others have been investigated in preclinical studies and clinical trials such as astatine-211 \big(\ce{^{211}At}\big) or actinium-225 \big(\ce{^{225}Ac}\big)\cite{eychenne2021}. Clinical evaluation of TAT has been performed for different types of cancer, including leukemia, lymphoma, melanoma, neuroendocrine tumors and various metastatic cancers\cite{Makvandi2018,poty2018b}. Among the major challenges faced by the scientific community for the translation of TAT to the routine clinical setting, one can mention the requirement of appropriate dosimetry approaches and imaging techniques to monitor treatment efficacy \cite{guerraliberal2020}.

Accurate dosimetry poses a major challenge in both BNCT and TAT because of the possible heterogeneous distribution of \ce{^{10}B} or the $\alpha$-emitter within the tumor. This issue may be specially problematic in BNCT because of the subcellular path length of the emitted radiations. Moreover, the underlying biological mechanisms triggered by low-energy ion irradiations are not yet fully understood. In particle therapy, biophysical models are often employed to estimate the RBE of ions for treatment optimization \cite{elsasser2010,friedrich2012,kase2006,sato2012,chen2017,carante2020,stewart2018,cunha2017a}. However, most of the existing biophysical models have been devised for high-energy ion beams, \eg for hadrontherapy applications. 
The translation of models for high-energy ion beams to BNCT and TAT is not straightforward as there is a need to consider the heterogeneity of dose deposition at the micro- and nanometer scale \cite{pedrosa2020simple, streitmatter2020mechanistic, horiguchi2015estimation}. A macroscopic dose is not sufficient to explain the observed biological effects. Several biophysical models are being developed to address this constraint, and some will be described in the Discussion. Moreover, RBE is a function of many parameters, like the microdistribution \cite{sato2018}, the deposited dose and the ion’s range, especially at the end of this range \cite{almhagen2023modelling}. This means that the energy lost by ions when traversing the biological targets should be taken into account to model the variations of LET, cell damage and RBE depending on the type of ion and its kinetic energy.  When this kind of biophysical model is not used, RBE is sometimes considered constant and equal to RBE$_{max}$ \cite{belli2020targeted, dale1999assessment}. However, RBE calculated for TAT and BNCT may range from 1 to 6, depending on the theoretical framework used \cite{ackerman2018targeted, mein2019biophysical, woollard2001development, sarnelli2022alpha}. Finally, Treatment Planning Systems (TPS) like Raystation \cite{kakino2022comprehensive} or NeuMANTA \cite{chen2022development} are including workflows for BNCT. They would benefit from those improved biophysical models. Our research team originally developed the NanOx (NANodosimetry and OXidative stress) \cite{cunha2017a} biophysical model for hadrontherapy applications. NanOx may be used to estimate the cell survival probability after exposure to ionizing radiations. NanOx calculations consider the stochasticity of energy deposits down to the nanometric scale, including the impact of the oxidative stress triggered by primary radical species. The general formalism of the NanOx model \cite{alcocer2023} and its implementation in hadrontherapy \cite{alcocer2022} were recently described in detail. 
The purpose of this paper is to show how the NanOx model can be adapted to irradiations with low-energy ions, which will be relevant for future applications in BNCT and TAT. The paper is structured as follows. We will first provide an overview of the general aspects of the NanOx model and introduce the approximations and simplifications required for applying the model to radiotherapies with low-energy ions. Then, we will describe the approach followed to validate these approximations and compare the hadrontherapy and low-energy implementations of NanOx. The positioning of NanOx within the wider literature on biophysical models and the perspectives for future improvements will be presented in the Discussion.


\section{Methods}
\label{sec:methods}
\label{sec:methods}
The general mathematical formalism of the NanOx model and its application to hadrontherapy have been described in detail in several papers \cite{alcocer2023,alcocer2022,cunha2017a}. For this reason, we provide in \Cref{sec:nanox_overview} only a brief summary of the principles and assumptions behind the model. The additional approximations required for implementing the model in the context of BNCT and TAT are introduced in \Cref{sec:approx_low_energy}. Two types of Monte Carlo (MC) simulations, developed for validating these approximations, are then presented in \Cref{sec:monte_carlo_applications}.

\subsection{Summary of the NanOx biophysical model}
\label{sec:nanox_overview}
NanOx is a biophysical model used to predict the cell survival probability following exposure to ionizing radiations. The model considers the stochastic nature of radiation interactions down to the nanometric scale, covering the physical, physico-chemical and chemical stages of radiation action by means of MC simulations \cite{gervais2006}. Radio-induced damage results from the energy deposited in the sensitive volume(s) of cells. In this work we will consider a single sensitive volume: the cell nucleus. Furthermore, cell survival depends on two types of events, assumed to be independent: the ``local'' and ``global'' lethal events.

Local events can be interpreted as complex DNA damage. They are evaluated in targets of nanometric dimensions uniformly distributed in the sensitive volume. The inactivation of one of such targets leads to cell death. This is described in NanOx by introducing the effective number of local lethal events (ENLLE), given by: \cite{alcocer2023,alcocer2022}:
\begin{linenomath*}
\begin{equation}
\Xk{\nE}{\ci,\ck}{} = - \ln \left( 1 - f(\Xk{z}{\ci,\ck}{}) \right)\ ,
\label{eq:meanENLLE}
\end{equation}
\end{linenomath*}

where the indices $\ci$ and $\ck$ represent the configuration of a local target $i$ in the sensitive volume and that of the radiation track $k$, respectively; $\Xk{f}{}{}(\Xk{z}{\ci,\ck}{})$ is the probability of inactivating the target $i$ when the radiation track $k$ deposits the restricted specific energy $\Xk{z}{\ci,\ck}{}$ in the target $i$. Let us recall that $z$, as defined within NanOx, includes only the processes and energy transfers that may be translated into biological effects. This is reflected through the ratio of the energy deposited by the considered processes to the total energy lost by radiation ($\eta\approx 0.8$) \cite{alcocer2023}.

The average ENLLE induced by the configuration $\ck$ of a radiation track $k$ in a sensitive volume containing $N$ targets with configuration $\cN$ may be written in an integral form as follows \cite{alcocer2023}:
\begin{linenomath*}
\begin{equation}
\Xk{\nE}{\cN,\ck}{} = \int_0^{+\infty} \prescript{\cN,\ck}{}{\left[\dv{P}{z}\right]} 
F(z) \dd{z}\ ,
\label{eq:cknintegral}
\end{equation}
\end{linenomath*}

where $\prescript{\cN,\ck}{}{\left[\dv{P}{z}\right]}$ is the density of probability for the radiation track configuration $\ck$ to deposit the restricted specific energy $\Xk{z}{\ci,\ck}{}$ in a local target $i$; and $F(z)$ is an effective local lethal function (ELLF) characterizing the response of a cell line \cite{monini2020}. In the following we will use the indices $\ti$ and $\tN$ to indicate the average response of a target $i$ and that of the $N$ local targets in the sensitive volume, respectively. The cell survival fraction to local events for a configuration $\cK$ of $K$ radiation tracks is then given by \cite{alcocer2022}:
\begin{linenomath*}
\begin{equation}
 \Xk{S}{\tN,\cK}{L} 
 =\exp (-\Xk{\nE}{\tN,\cK}{})\ .
\label{eq:cellSurv_ENLLE}
\end{equation}
\end{linenomath*}

On the other hand, global events result from the accumulation of sublethal lesions and the oxidative stress produced by free radicals. For a radiation track $k$ with configuration $\ck$, the concentration of primary reactive chemical species is given by\cite{alcocer2023}:
\begin{equation} 
\Xk{Y}{\ck}{} = \frac{\Gr}{\eta}\cdot \Xk{\Zchem}{\ck}{} = \frac{\Xk{G}{\ck}{}}{\eta}\cdot \Xk{Z}{\ck}{}\ ,
\label{eq:Ydef} 
\end{equation}

where $\Gr$, $\Xk{G}{\ck}{}$ are the chemical yields for the reference radiation and the radiation of interest, respectively; and $\Xk{\Zchem}{\ck}{}$ is called the chemical specific energy. The cell survival fraction to global events for a configuration $\cK$ of radiation tracks is given by\cite{alcocer2023}:
\begin{linenomath*}
\begin{equation}
\Xk{S}{\cK}{G}  =
     \exp\left( -\aG \Xk{\Zchem}{\cK}{} -\bG \Xk{\Zchem}{\cK}{}^2 \right)\ ,
\label{eq:cellSurv_globalEv_Ztilde}
\end{equation}
\end{linenomath*}

where $\aG$ and $\bG$ are characteristic of the cell line and are obtained from cell survival curves for reference radiation.
\sisetup{per-mode=reciprocal}

\subsection{Extension of the NanOx model to low-energy ion irradiations}
\label{sec:approx_low_energy}
In this section, we present the approximations and simplifications considered in NanOx to model irradiations with low-energy ions.
\subsubsection{Characterization of the radiation tracks and the sensitive volume}
\label{sec:Vs}
In hadrontherapy, the high energy of ion beams allows the assumption of track-segment conditions, i.e. the primary projectile moves in a straight line and its energy (and LET) remains constant while crossing the target. By contrast, BNCT and TAT use low-energy, short-range ions which may lose a great amount of energy while traversing micrometric volumes, \eg subcellular structures. In consequence, the modeling of low-energy ions has to consider the evolution of the ion's energy and trajectory across the sensitive volume.

\approximation{Narrow radiation tracks}{hyp:narrow_tracks}{
The energy of the ions is low enough to consider the radiation tracks as narrow. The latter implies that the energy deposited by the secondary electrons outside the sensitive volume that the ion is traversing is assumed to be negligible. We shall consider in the following that this approximation holds whenever the proportion of energy deposited by secondary electrons beyond a radius of 100 nm around the ion's track is $\lesssim$ 1\% (see \Cref{sec:valid_nt,sec:result_nt}).
}
\approximation{Charge equilibrium}{approx:boundary_effects}{
The energy deposited in the sensitive volume is equal to the difference of the ion's kinetic energies at the entry ($E_{\mathrm{i}}$) and exit ($E_{\mathrm{f}}$) points:
\begin{linenomath*}
\begin{equation}
\Delta E_{\mathrm{dep}} = E_{\mathrm{i}} - E_{\mathrm{f}}\ .
\label{eq:edep}
\end{equation}
\end{linenomath*}
}
As it is demonstrated later, the aim of this approximation is to greatly reduce computation time and storage by working with only two energy values instead of hundreds of thousands of energy transfers describing the tracks. The limit of this approximation is directly related to the equilibrium of charged particles. For low-energy ions, the depth of equilibrium is less than 10 nm, which is small compared with the cell geometry.

\simplification{Same sensitive volume for local and global events}{simpl:local_nonLocal_SV}{
As in hadrontherapy\cite{alcocer2022}, we assume that local and global lethal events take place in the same sensitive volume, the cell nucleus. 
}

\subsubsection{Calculation of local and global lethal events for low-energy ions}
\label{sec:calc_lle_ge}

\paragraph{Local lethal events}\mbox{}
\label{sec:lle_low_energy}

\begin{figure}[h]
\centering
\includegraphics[width=8cm]{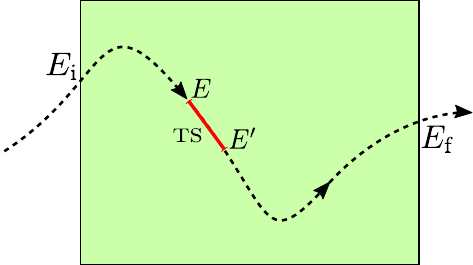}
\caption{Schematic representation of the trajectory of an ion entering and leaving the sensitive volume. The track segment TS satisfying the conditions listed in the text is represented by the solid red line.\\}
\label{fig:trajectory}
\end{figure}

Let us now consider an ion track of configuration $\ck$ and a track segment TS of length $\left(\Delta \ell\right)_{\mathrm{TS}}$ along which the ion moves in a straight line and loses the energy $\Delta\Xk{E}{\ck}{}$ (see \Cref{fig:trajectory}):
\begin{equation}
\Delta\Xk{E}{\ck}{} = E - E'\ ,
\label{eq:deltaE_TS}    
\end{equation}

with $E$ and $E'$ the energy of the ion at the beginning and end of the track segment TS, respectively.
\FloatBarrier
\FloatBarrier

In addition, we consider that the following conditions hold.
\condition{Small energy loss along $\left(\Delta \ell\right)_{\mathrm{TS}}$}{cond:deltaE}{
The energy loss $\Delta\Xk{E}{\ck}{}$ along $\left(\Delta \ell\right)_{\mathrm{TS}}$ is small enough so that:
\begin{linenomath*}
\begin{equation}
\frac{\Delta\Xk{E}{\ck}{}}{E} = \varepsilon\ , \quad \varepsilon \ll 1
\label{deltaEE}    
\end{equation}
\end{linenomath*}

\condition{Charged-particle equilibrium}{cond:parteq}{
$\left(\Delta \ell\right)_{\mathrm{TS}}$ is larger than the range $R$ of the $\delta$-rays:
\begin{linenomath*}
\begin{equation}
\left(\Delta \ell\right)_{\mathrm{TS}} > R\ .
\label{eq:deltalrays}    
\end{equation}
\end{linenomath*}
}

\condition{Track segment much larger than the nanotarget length}{cond:relsizesegment}{
$\left(\Delta \ell\right)_{\mathrm{TS}}$ is much larger than the nanotarget length, that is:
\begin{linenomath*}
\begin{equation}
\left(\Delta \ell\right)_{\mathrm{TS}} \gg L_{\mathrm{t}}\ ,
\label{eq:deltalnano}    
\end{equation}
\end{linenomath*}

where $L_{\mathrm{t}}$ is the length of the nanotargets considered in the calculations (10 nm).
}
We can then write $\left(\Delta \ell\right)_{\mathrm{TS}}$ as:
\begin{linenomath*}
\begin{equation}
\left(\Delta \ell\right)_{\mathrm{TS}} = \Delta\Xk{E}{\ck}{} \cdot \left(\dv{E}{x}\right)^{-1} = \varepsilon\cdot E\cdot\left(\dv{E}{x}\right)^{-1} , 
\label{eq:deltal}    
\end{equation}
\end{linenomath*}
}

Moreover, we can then associate a surface of influence $\Sigma_{\mathrm{TS}}$, a volume $\left(\Delta V\right)_{\mathrm{TS}}$ and a mass $\left(\Delta m\right)_{\mathrm{TS}}$ to the track segment, such that:
\begin{equation}
\left(\Delta V\right)_{\mathrm{TS}} = \Sigma_{\mathrm{TS}}\cdot \left(\Delta \ell\right)_{\mathrm{TS}} = \frac{\left(\Delta m\right)_{\mathrm{TS}}}{\rho_{\mathrm{w}}}\ ,
\label{eq:surfTS}    
\end{equation}

with $\rho_{\mathrm{w}}$ the density of water. The surface of influence $\Sigma_{\mathrm{TS}}$ is chosen to be large enough so that any ion generated outside the volume of influence will have no chance of depositing energy in the sensitive volume. We then define the ENLLE for an ion track $\ck$ as:
\begin{equation} 
\Xk{n^*}{\tN, \ck}{}=~\sum_{\mathrm{TS}}{\left(\Delta\Xk{n^*}{\tN, \ck}{}\right)_{\mathrm{TS}}}  
\ ,
\label{eq:enlleTS}
\end{equation}

Introducing the energy loss $\Delta\Xk{E}{\ck}{}$, we have:
\begin{equation}
\left(\Delta\Xk{n^*}{\tN, \ck}{}\right)_{\mathrm{TS}} =  
\prescript{\tN,\ck}{}{\left(\frac{\Delta n^*}{\Delta E}\right)_{\mathrm{TS}}}
\Delta\Xk{E}{\ck}{}\ ,
\label{eq:deltanz}    
\end{equation}

Where the term $\prescript{\tN,\ck}{}{\left(\frac{\Delta n^*}{\Delta E}\right)_{\mathrm{TS}}}$ is given by:
\begin{equation}
\prescript{\tN,\ck}{}{\left(\frac{\Delta n^*}{\Delta E}\right)_{\mathrm{TS}}} =
\frac{\left(\Delta V\right)_{\mathrm{TS}}}{V_{\mathrm{s}}}\cdot \frac{1}{ \Delta\Xk{E}{\ck}{}}
\int_{0}^{+\infty} \prescript{\ti,\ck}{}{\left[\dv{P}{z} \right]_{\mathrm{TS}}}   
\Xk{F(z)}{\tN}{} \dd z\ ,
\label{eq:intFz_ck}
\end{equation}

where $V_{\mathrm{s}}$ is the sensitive volume. Furthermore, as the track is narrow, we can apply the same approximation made for the core volume of an ion's track in the application of the NanOx model to hadrontherapy \cite{alcocer2022}, but for a track segment TS.

\approximation{Negligible fluctuations from a configuration $\ck$ to another}{approx:cktotk}{
The fluctuations in the term $\prescript{\tN,\ck}{}{\left(\frac{\Delta n^*}{\Delta E}\right)_{\mathrm{TS}}}$ may be seen as negligible from a configuration $\ck$ to another, so it is possible to take the average over a large number of particles of the same type $t_{\mathrm{k}}$ and energy $E_{\mathrm{k}}$. We use the index $t_{\mathrm{k}}$ to denote this approximation:
\begin{equation} 
\prescript{\tN,\ck}{}{\left(\frac{\Delta n^*}{\Delta E}\right)_{\mathrm{TS}}} = 
\prescript{\tN,\tk}{}{\left(\frac{\Delta n^*}{\Delta E}\right)_{\mathrm{TS}}} \ .
\label{eq:ncktotk}
\end{equation} 
}
And thus we can rewrite \Cref{eq:intFz_ck}:
\begin{equation}
\prescript{\tN,\tk}{}{\left(\frac{\Delta n^*}{\Delta E}\right)_{\mathrm{TS}}} =
\frac{\left(\Delta V\right)_{\mathrm{TS}}}{V_{\mathrm{s}}}\cdot \frac{1}{\Delta\Xk{E}{\ck}{}} 
\int_{0}^{+\infty} \prescript{\ti,\tk}{}{\left[\dv{P}{z} \right]_{\mathrm{TS}}}  \Xk{F(z)}{\tN}{} \dd z\ ,
\label{eq:intFz_tk}
\end{equation}

At this point we will consider for practical purposes a mesoscopic scale between the nanometric scale of the targets and the microscopic scale of the radiation tracks. The latter hypothesis will allow us to work with a differential formulation of the problem:
\begin{equation}
\prescript{\tN,\tk}{}{\left(\frac{\Delta n^*}{\Delta E}\right)_{\mathrm{TS}}} =
\prescript{\tN,\tk}{}{\left(\frac{\Xk{\dd n^*}{}{}}{\Xk{\dd E}{}{}}\right)}\ .
\label{eq:dndE}    
\end{equation}

Then, ignoring any boundary effect, we can write:
\begin{equation} 
\Xk{n^*}{\tN, \tk}{} = \int^{\Xk{E_{\mathrm{f}}}{\tk}{}}_{\Xk{E_{\mathrm{i}}}{\tk}{}} 
{\prescript{\tN,\tk}{}{\left(\frac{\Xk{\dd n^*}{}{}}{\Xk{\dd E}{}{}}\right)} (-\dd E)} = 
\int^{\Xk{E_{\mathrm{i}}}{\tk}{}}_{\Xk{E_{\mathrm{f}}}{\tk}{}} 
{\prescript{\tN,\tk}{}{\left(\frac{\Xk{\dd n^*}{}{}}{\Xk{\dd E}{}{}}\right)} \dd E} =
\phi(\Xk{E_{\mathrm{i}}}{\tk}{}) -\phi(\Xk{E_{\mathrm{f}}}{\tk}{})\ , 
\label{eq:tNtkn_integral}
\end{equation}

where $\phi$ is a primitive of $\left(\frac{\Xk{\dd n^*}{}{}}{\Xk{\dd E}{}{}}\right)$ and $\Xk{E_{\mathrm{i}}}{\tk}{}$, $\Xk{E_{\mathrm{f}}}{\tk}{}$ denote the energy of the ion at the beginning and end of the track in the sensitive volume, respectively. The beginning of this track may correspond to the entry of the ion into the sensitive volume or to the creation of this ion within the sensitive volume. The end of the track could be the point at which the ion leaves the sensitive volume. Alternatively, it may correspond to the full stopping of the ion ($\Xk{E_{\mathrm{f}}}{\tk}{} = 0$), or to the production of a nuclear process. Moreover, for each track there may be several entries and exits. 

\subsubsection{Global events}
\label{sec:ge_low_energy}
For global events, it is also possible to take benefit of the narrow-track approximation. The concentration of primary reactive chemical species may be written as (from \Cref{eq:Ydef}):
\begin{equation} 
\Xk{Y}{\tk}{} = \frac{1}{m_{\mathrm{s}}} \int^{\Xk{E_{\mathrm{f}}}{\tk}{}}_{\Xk{E_{\mathrm{i}}}{\tk}{}} {\Xk{G(E)}{\tk}{} (-\dd E)} =
\frac{1}{m_{\mathrm{s}}}
\int^{\Xk{E_{\mathrm{i}}}{\tk}{}}_{\Xk{E_{\mathrm{f}}}{\tk}{}} {\Xk{G(E)}{\tk}{} \dd E} = \frac{\Gr}{\eta}\cdot \Xk{\Zchem}{\tk}{}\ ,
\label{eq:Yintegral} 
\end{equation} 
with $m_{\mathrm{s}}$ the sensitive volume's mass and:
\begin{equation} 
\Xk{\Zchem}{\tk}{} =
\frac{\eta}{m_{\mathrm{s}}}
\left[\psi\left(\Xk{E_{\mathrm{i}}}{\tk}{}\right) - 
\psi\left(\Xk{E_{\mathrm{f}}}{\tk}{}\right)\right]\ ,
\label{eq:Zchem_primitives} 
\end{equation} 
with $\psi$ a primitive in $E$ of $\frac{\Xk{G}{\tk}{}}{\Gr}$.

It is worth emphasizing that in this paper all calculations regarding the low-energy NanOx implementation were based on the integration of local lethal events and global events (yields of reactive chemical species) along the ions' path in the sensitive volume (\Cref{eq:tNtkn_integral,eq:Zchem_primitives}).

\subsection{Validation of approximations and test of the low-energy NanOx implementation}
\label{sec:monte_carlo_applications}

\subsubsection{Validation of the narrow-track approximation}
\label{sec:valid_nt}
To verify the validity of the narrow-track approximation, the proportion of energy deposited outside the core volume of the ion's track  was calculated by means of MC simulations. The core volume is the region marked by a high density of energy-transfer points around the ion's track \cite{alcocer2022}. In this work, we modeled the core volume as a cylinder of radius 100, 200 or 300~nm centered around the ion's track. Helium ions of 7.5 MeV (maximum energy of the $\alpha$-particles emitted by \ce{^{211}At}\big) were shot in a water cube, big enough for the ions to fully stop in it. The energy deposited inside and outside the cylinder was then tallied. These MC simulations were performed with Geant4-DNA \cite{incerti2010}, using the option 2 physics constructor.

\subsubsection{Validation of the charge equilibrium approximation}
Approximation \ref{approx:boundary_effects} (\Cref{eq:edep}) was also validated through MC simulations. The Geant4 \cite{allison2016} simulation toolkit, version 11.1, was used with the standard electromagnetic and DNA \cite{incerti2010} physics lists. The objective was to calculate the relative difference between $(E_{\mathrm{i}} - E_{\mathrm{f}})$ and $\Delta E_{\mathrm{dep}}$ in respect of their average:
\begin{equation}
\sigma_E = 2\cdot \frac{(E_{\mathrm{i}} - E_{\mathrm{f}}) - \Delta E_{\mathrm{dep}}}{(E_{\mathrm{i}} - E_{\mathrm{f}}) + \Delta E_{\mathrm{dep}}}\ .
\label{eq:rel_diff_ei_ef_edep}
\end{equation}

The irradiation with $\alpha$-particles was simulated with Geant4, mimicking a TAT application. A cell geometry consisting of 2 concentric spheres with a cell radius of 7.1 µm and nucleus radius of 5.2 µm was considered to model the human salivary gland (HSG) cell line \cite{choi2014trpv1}. $\alpha$-particles were generated isotropically in the cell cytoplasm, with the initial kinetic energy varying from 0.1 to 50 MeV/n. The ions were tracked until their energy reached 1 keV. For each ion track, $E_{\mathrm{i}}$, $E_{\mathrm{f}}$ and $\Delta E_{\mathrm{dep}}$ were collected. For each configuration, 10$^8$ ions were sent. The errors were calculated as the standard deviation divided by $\sqrt{N}$, with $N$ the number of events. 

\subsubsection{Comparison of predictions using the hadrontherapy and low-energy implementations of NanOx}
\label{sec:formalism_comparison}
\begin{figure}[ht]
\centering
\vspace{3mm}
\includegraphics[height=8cm]{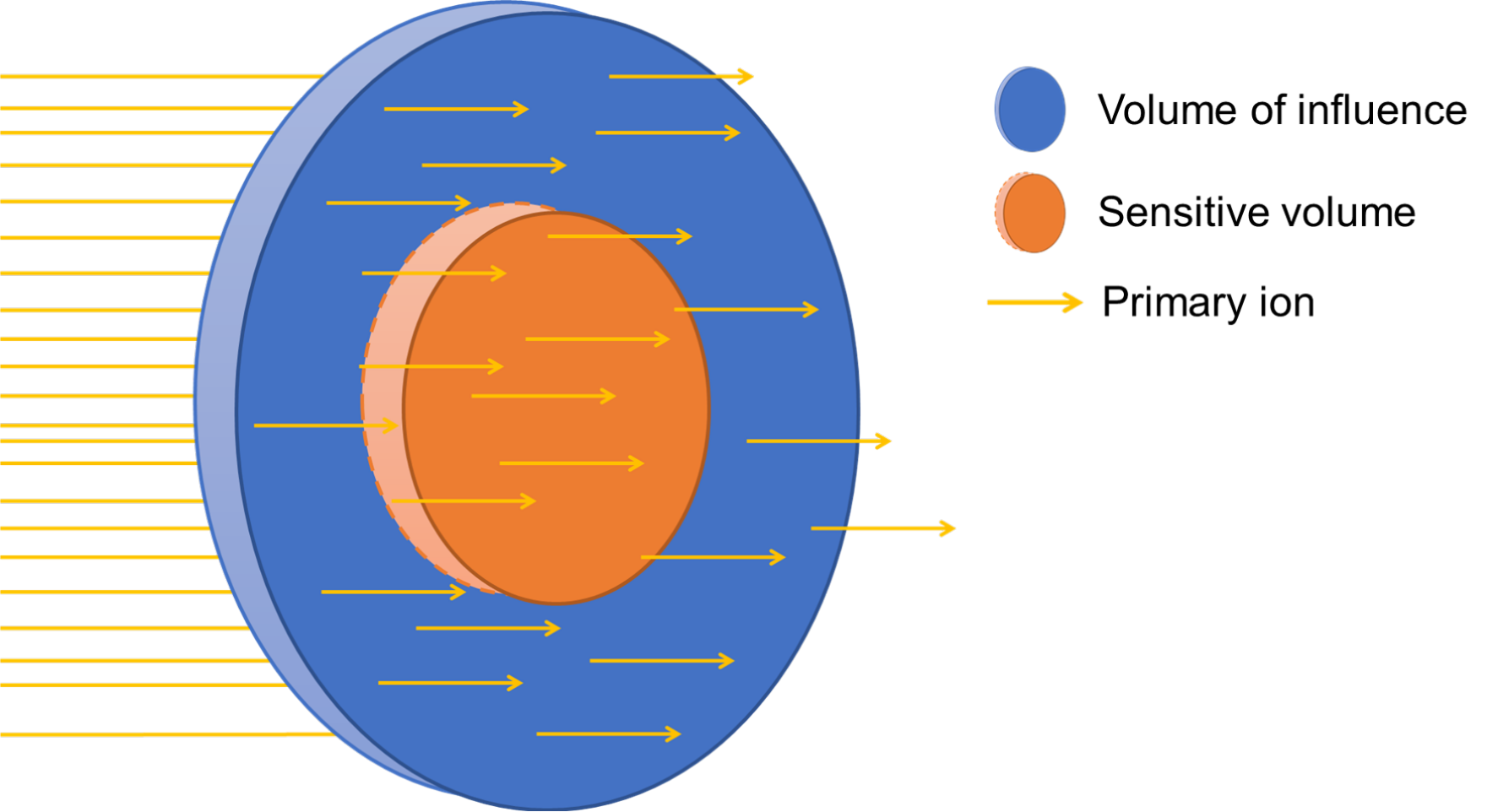}
\caption{Geometry used in the simulations to calculate the inactivation cross sections. The inner cylinder is the sensitive volume, namely the cell nucleus; the outer cylinder is the volume of influence. The straight lines represent $\alpha$-particles with their secondary electrons.}
\label{fig:geom_sigma_simulation}
\end{figure}

To compare the NanOx implementation described in this paper and the one applied in hadrontherapy \cite{alcocer2022}, the chosen quantity was the inactivation cross section $\sigma$. It corresponds to the efficiency of an ion to kill a cell. It is calculated as \cite{alcocer2023}:
\begin{equation}
    \Xk{\sigma}{\tk}{} = \Sigma_{\mathrm{TS}} \cdot (1 - \Xk{\langle S_1 \rangle}{\tk}{})\ ,
\label{eq:sigmaInactivationCrossSec}
\end{equation}

where $\Sigma_{\mathrm{TS}}$ denotes the chosen surface of influence; $\Xk{\langle S_1 \rangle}{\tk}{}$ is the average survival when one particle is sent to the volume of influence, computed over a large number of configurations. We remind that, in this study, the sensitive volume is the cell nucleus. MC simulations were performed to obtain the results for the low-energy NanOx implementation, using the geometry presented in figure \ref{fig:geom_sigma_simulation} and the standard electromagnetic physics list of Geant4 (option 4).
The sensitive volume and the volume of influence consisted of coaxial cylinders of equal length, measured along the z-axis. Different simulations were performed to evaluate the effect of modifying the length of the cylinders. The number of nanometric targets in the sensitive volume remained constant, regardless of its length. This choice was made for facilitating the comparison between the inactivation cross section curves in a wide energy range. The geometry was chosen to reproduce the MC simulation conditions of NanOx, depicted in figure 2 of Alcocer-Ávila et al. (2022) \cite{alcocer2022}. The radius of 7 µm for the sensitive volume was chosen to match the reference value of the HSG nucleus radius reported in Monini et al. (2019)\cite{monini2019}. The radius of the volume of influence was 14 µm. All ions were sent at the entrance of the volume of influence. If the sensitive volume was hit, the cell survival was calculated. Otherwise, its value was equal to 1. The parameters of the ELLF, $F(z)$, and the values of $\aG$ and $\bG$ (\Cref{eq:cellSurv_globalEv_Ztilde}) are those formerly reported for the HSG cell line \cite{alcocer2022}. For the hadrontherapy application the results were extracted from our previous studies \cite{alcocer2022, monini2019}.

The NanOx model does not take into account elastic nuclear interactions. At very low kinetic energies, the nuclear stopping power of $\alpha$-particles in water may not be negligible compared to electronic stopping power, according to the NIST ASTAR database \cite{astar2005}. Furthermore, under 100~keV/n the model does not provide the density of local and global lethal events per energy. Without experimental data to evaluate the impact of such low energies, hypotheses have to be made. Three will be compared. First, the $\alpha$ coefficient was considered equal to zero below 100~keV/n. Second, $\alpha$ was set to zero at a kinetic energy equal to zero, and a linear interpolation was made between zero and the last known value of $\alpha$, i.e. the one at 100 keV/n. Third, $\alpha$ was considered constant and equal to the value at 100 keV/n, for all energies below 100~keV/n.


\section{Results}
\label{sec:results}
In this section, we report the results of our MC simulations performed to validate the narrow-track and charge equilibrium approximations used in the low-energy NanOx implementation. In addition, we compare the predictions of the hadrontherapy and low-energy NanOx implementations, showing that there exists a region of intermediate energies in which both implementations overlap, leading to the same predictions.

\subsection{Validation of the narrow-track approximation}
\label{sec:result_nt}
When the core volume was modeled as a cylinder of 100, 200 and 300~nm radius, the proportion of energy deposited outside of it was 1.28\%, 0.22\% and 0.05\%, respectively. These values are averaged over the whole ion's track. Table \ref{tab:NarrowTrackApprox_WithCoreVolume} presents the results obtained for the core volume of 100~nm radius, when considering 10~µm slices along the full path of the helium ion ($\sim$ 70 µm).\\
\FloatBarrier
\begin{table}[h]
\centering
\begin{tabular}{lc}
\toprule
\begin{tabular}[c]{@{}l@{}} Axial position on\\ the ion's track \end{tabular} &
\begin{tabular}[c]{@{}l@{}} Energy deposit outside \\ the core volume \end{tabular} \\
\cmidrule(r){1-1} \cmidrule(rl){2-2}
0 - 10~µm  & 4.33\% \\ 
10 - 20~µm & 3.32\% \\ 
20 - 30~µm & 0.76\% \\ 
30 - 40~µm & 0.41\% \\ 
40 - 50~µm & 0.06\% \\
50 - 60~µm & 0.0\%  \\
60 - 70~µm & 0.0\%  \\
\bottomrule
\end{tabular}
\caption{Proportion of energy deposited outside a cylinder of 100~nm radius representing the core volume of a 7.5 MeV helium ion track. The ion's path was divided in slices of 10~µm to explore non-averaged effects.\\}
\label{tab:NarrowTrackApprox_WithCoreVolume}
\end{table}
\FloatBarrier
For the cylinders of 200 and 300~nm radius, the proportion of energy deposited outside the core volume was always below 1\%.

\subsection{Validation of the charge equilibrium approximation} 
\begin{figure}[ht]
\centering
\includegraphics[height=8 cm]{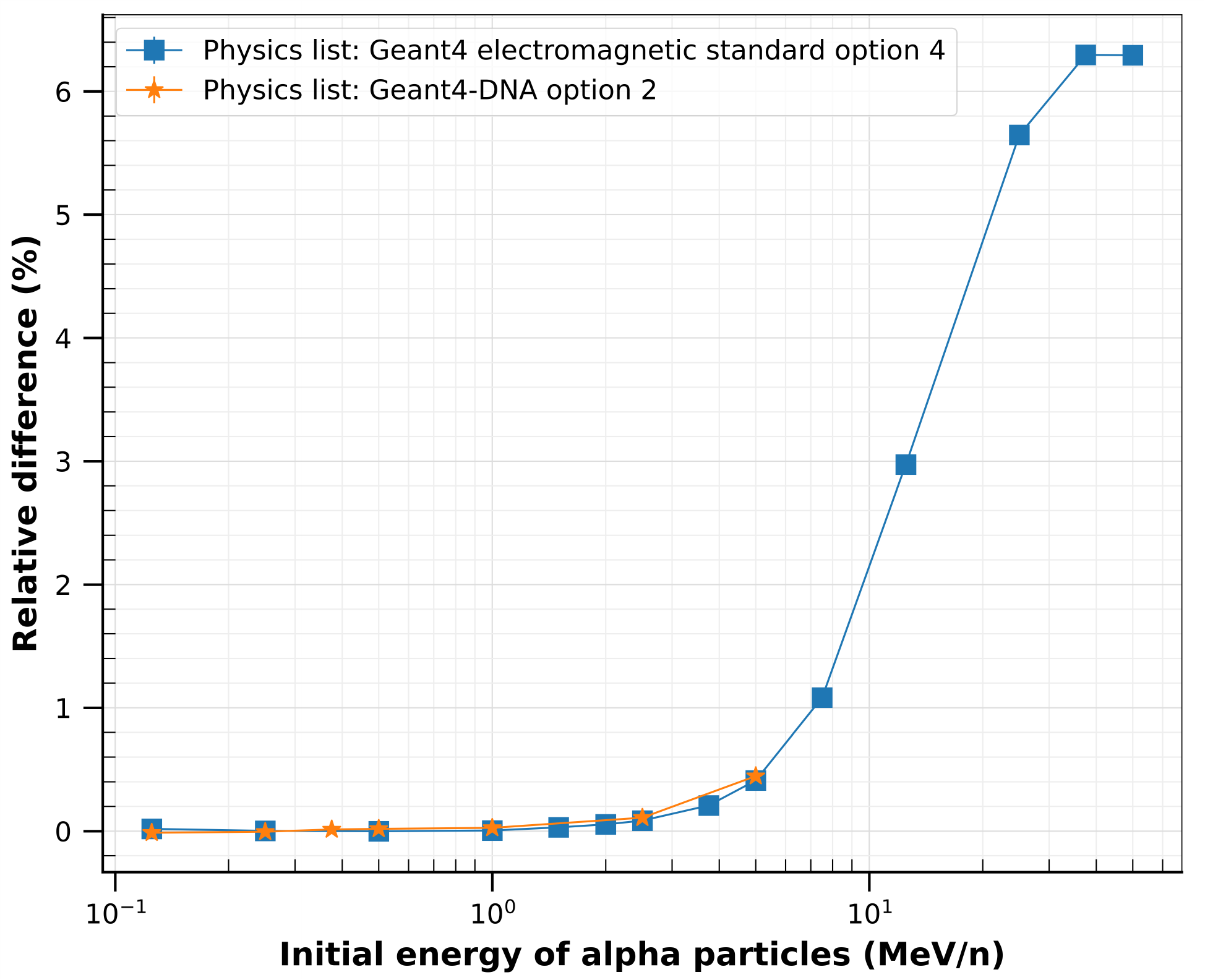}
\caption{The relative difference $\sigma_E$ (\Cref{eq:rel_diff_ei_ef_edep}) is shown as a function of the initial kinetic energy of $\alpha$-particles in MeV/n. Ions are emitted isotropically from the cytoplasm of spherical cells. The solid lines are plotted for visualization purposes only. Statistical errors are contained within the data points.\\}
\label{fig:ei_ef_verification}
\end{figure}

We present in this section the results that allowed us to validate the charge equilibrium approximation. Figure \ref{fig:ei_ef_verification} presents the relative difference $\sigma_E$ as defined in \Cref{eq:rel_diff_ei_ef_edep}, expressed in percentage as a function of the initial kinetic energy of $\alpha$-particles.
\FloatBarrier
The standard electromagnetic physics list option 4 is compared to the Geant4-DNA option 2 physics list. The first one is the most precise standard physics list at low energy, and all generated electrons were tracked down to 1 keV. The second one has a more precise tracking of electrons in water, tracked down to about 7 eV. Below 1 MeV/n, the relative difference observed is approximately equal to 0. Differences of 1\% show up at 8 MeV/n. They continue to increase up to 40 MeV/n, where there are differences of ~6\%. No important differences were noted between the two physics lists.

\subsection{Comparison of predictions using the hadrontherapy and low-energy implementations of NanOx}
\label{subsec:inactivationcrosssection_results}
\begin{figure}[ht!]
  \centering
  \begin{subfigure}{\textwidth}
    \centering
    \includegraphics[height=8cm]{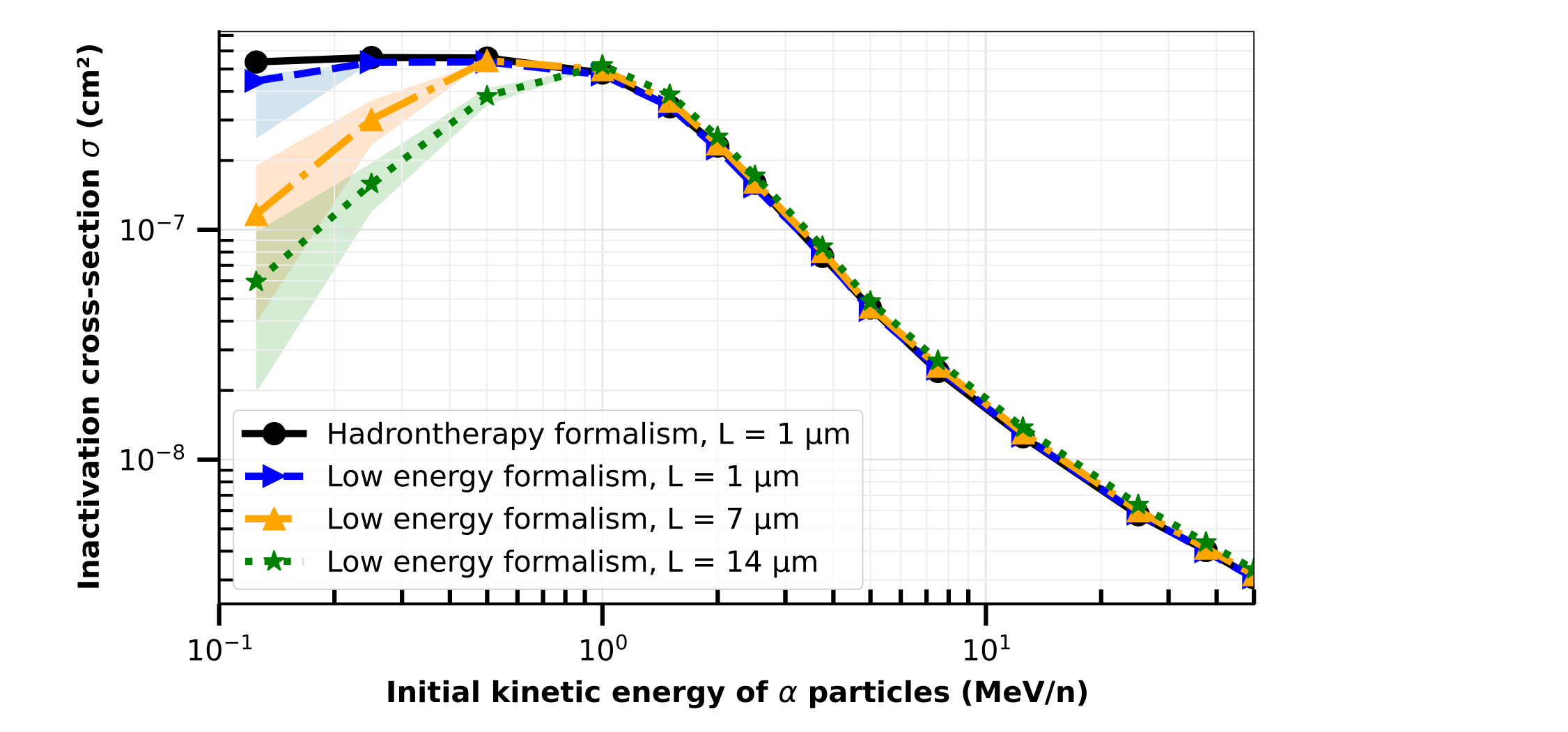}
    \caption{}
    \label{subfig:sigma_comparison}
  \end{subfigure}
  \begin{subfigure}{\textwidth}
    \centering
    \includegraphics[height=8cm]{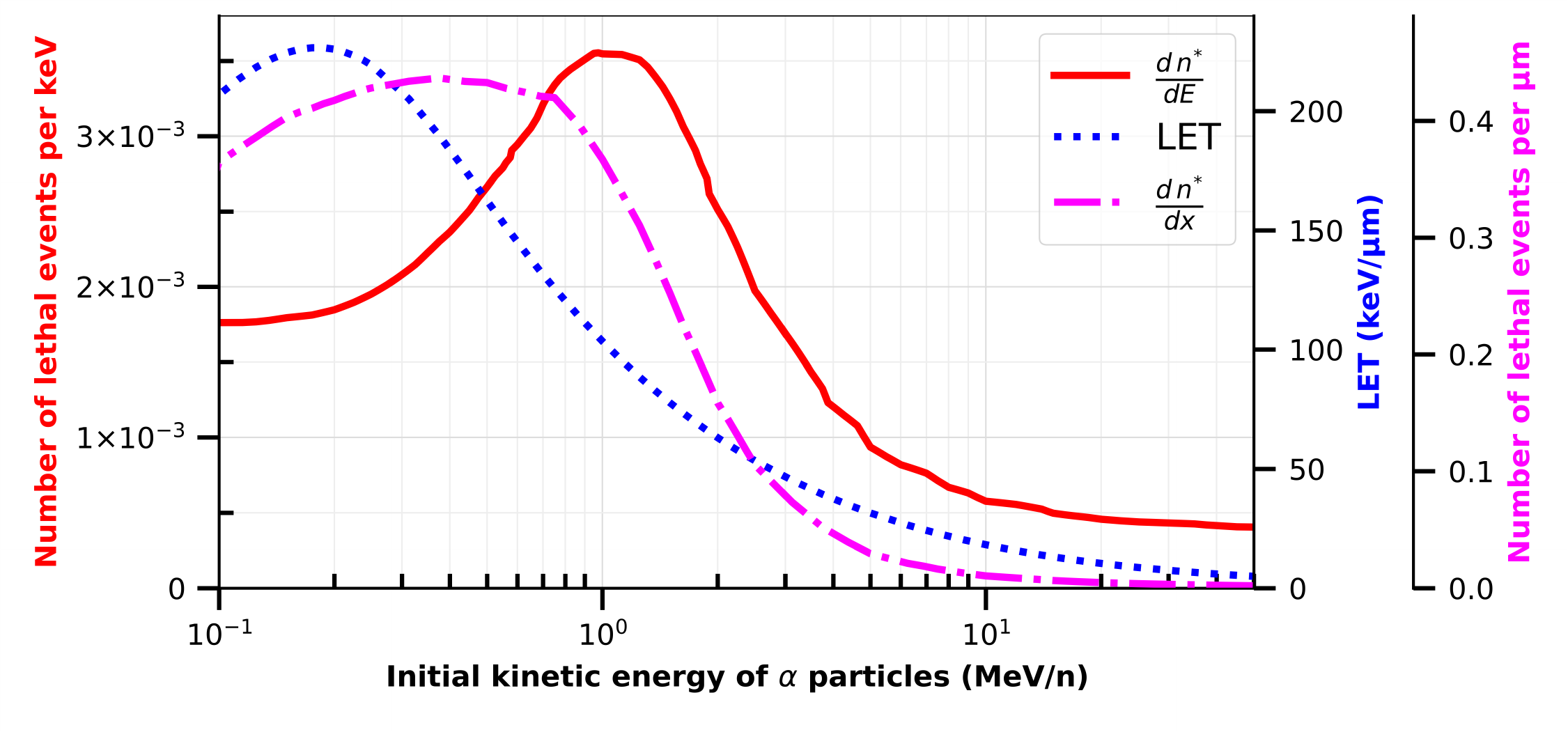}
    \caption{}
    \label{subfig:dn_de_let}
  \end{subfigure}
  \caption{\textbf{(a)}: Inactivation cross section (cm²) as a function of the initial kinetic energy of $\alpha$-particles (MeV/n). The solid circles are the results obtained with the hadrontherapy implementation. The results of the low-energy implementation are represented by the right-pointing triangles, the up-pointing triangles, and the stars for the 1, 7, and 14 µm target thicknesses, respectively. The shaded areas correspond to the range of variation of $\sigma$ for the three previously described hypotheses about the behavior of the $\alpha$ coefficient below 100 keV/n (see \Cref{sec:formalism_comparison}). \textbf{(b)}: number of lethal events per keV (solid line, left axis), LET (dotted line, right axis), and number of lethal events per µm (dash-dotted line, right axis) as a function of the incident kinetic energy of $\alpha$-particles. The number of lethal events per µm is the product of number of lethal events per keV and the LET.}
  \label{fig:sigma_comparison_and_dn_de_let}
\end{figure}

 Figure \ref{subfig:sigma_comparison} represents the inactivation cross section (in cm²) as a function of the initial kinetic energy of $\alpha$-particles (MeV/n). The length of the cylindrical targets was varied between 1 and 14 µm in the low-energy NanOx implementation, 14 µm corresponding to the mean diameter of the nucleus for the HSG cell line. In the hadrontherapy implementation, the energy lost by ions is not considered, hence the length of the target has no influence.
 
\FloatBarrier
Good agreement is observed in the inactivation cross sections ($\sigma$) computed with the two implementations of NanOx for kinetic energies between 1 and 50 MeV/n, irrespective of the target geometry.\\
Below 1 MeV/n, the low-energy implementation results in smaller $\sigma$ values. These differences are visible below 0.25, 0.5 and 1 MeV/n for the 1, 7 and 14 µm target thicknesses, respectively. Differences become larger as the kinetic energy of the $\alpha$-particles decreases. The maximum difference with respect to the results obtained with the hadrontherapy implementation was found at 0.125 MeV/n for the 14~µm length cylinder, with a 9-fold smaller cross section.\\
The choice of the hypothesis to describe the behavior of the $\alpha$ coefficient below 100 keV/n had no influence on the inactivation cross sections for kinetic energies between 1 and 50~MeV/n. For 0.125 MeV/n, changing the hypothesis leads to $\pm$ 65\% difference in $\sigma$, for the 7 and 14 µm thick targets (shaded areas in \Cref{subfig:sigma_comparison}). For the target of 1 µm, the hypothesis that lowered the most the $\alpha$ coefficient led to a $-44$\% decrease in $\sigma$. The higher the kinetic energy of $\alpha$-particles, the lower the impact of the hypothesis on the inactivation cross section.\\
Figure \ref{subfig:dn_de_let} shows, on the left axis, the density of lethal events per keV, $\frac{dn^*}{dE}$, for $\alpha$-particles calculated with NanOx. On the right axis is the LET in keV/µm, with data taken from the NIST ASTAR database \cite{astar2005}, and the number of lethal events per µm, $\frac{dn^*}{dx}$. All are plotted against the initial kinetic energy of $\alpha$-particles.\\
The density of lethal events per keV and the LET exhibit a similar trend as a function of energy, but the peak of the LET curve is shifted towards lower energies: the maximum density of lethal events per keV  occurs at 1 MeV/n; in contrast, the maximum LET is found at approximately 0.18 MeV/n. Below 1 MeV/n, the density of lethal events per µm is between 0.35 and 0.43 µm$^{-1}$.\\
Finally, let us note that when we consider the case of intermediate-energy ions, it is possible to find an expression relating directly $\frac{\Xk{\dd n^*}{}{}}{\Xk{\dd E}{}{}}$ and the radiobiological $\alpha$ coefficient (see \Cref{sec:appendix_dnde_alpha}).


\section{Discussion}
\label{sec:discussion}
\subsection{Narrow-track and charge equilibrium approximations}
To study the biological effects of low-energy ions, precision at the micrometric scale is needed. Previous applications of the NanOx model considered the track-segment approximation: no energy was lost by ions when traversing the sensitive volume (i.e. the cell nucleus). As the energy of ions decreases, alternative assumptions must be made to accurately describe the energy deposition at the cellular scale. This is the rationale for introducing a low-energy NanOx implementation in this paper. To simplify calculations, the narrow-track and the charge equilibrium approximations were introduced (\Cref{sec:Vs}).\\ 
The verification of the first approximation was made by mimicking in the MC simulations the separation of an ion's track in core and penumbra volumes, as usually considered in NanOx \cite{alcocer2023}. At the beginning of an ion's track, the emitted secondary electrons are more energetic than at the end. These energetic electrons may easily escape the core volume, hence there is more energy deposited outside the core volume at the beginning of the track (see \Cref{tab:NarrowTrackApprox_WithCoreVolume}). The energy of the $\alpha$-particles considered for this test was the highest that can be encountered in TAT or BNCT, which represents the case with the greatest probability of secondary electrons escaping from the core volume. Overall, however, we have shown that the proportion of energy deposited in the penumbra volume remains small for low-energy ions. In this case, the use of the narrow-track approximation is justified and calculations are simplified by removing the distinction between core and penumbra.\\
The charge equilibrium approximation is linked with the previous one: they are both due to the relatively short range of the secondary electrons produced by low-energy ions. The validation of the resulting approximation (\Cref{eq:edep}) was made with a distribution of $\alpha$-particles that is likely to be found in TAT, which is not favorable for our approximation. The closer the $\alpha$-particles are emitted to the nucleus, the greater the probability that their secondary electrons will escape from it. However, the approximation considers that the generated secondary electrons deposit all their energy in the nucleus. In TAT, the energy of $\alpha$-particles will never be higher than 2.5 MeV/n \cite{tronchin2022dosimetry}. In our simulation study, the relative difference between the deposited energy and the ion's kinetic energy variation in target (\Cref{eq:rel_diff_ei_ef_edep}) was $<$ 0.1\% for kinetic energies below 2.5 MeV/n. This validates the use of the approximation in this paper for targeted radiotherapies involving low-energy ions. Therefore, electrons do not need to be precisely tracked in the simulations. Thanks to this approximation, the calculation time of our Geant4 simulations was divided by 3, and the storage needed for the output files was divided by more than a million.

\subsection{Application of NanOx to low-energy ions}
The agreement of the low-energy NanOx implementation with the hadrontherapy one is surprising. On the one hand, between 1 and 50 MeV/n, no important difference can be found between the two implementations of the model. The differences could have come from energetic, long-range $\delta$ electrons that are considered in the hadrontherapy implementation and not within the narrow-track approximation of this paper. However, in the energy range considered here, $\delta$ electrons do not contribute enough to energy deposits that could induce biological effects. Above 10~MeV/n, the energy lost by an ion in a nucleus is negligible compared to its initial kinetic energy (see \Cref{sec:appendix_eief}). The low-energy implementation has no bearing in this case because the micrometric precision in the description of the energy deposition process is no longer needed. These high energies are beyond the scope of the low-energy NanOx implementation presented in this paper, so no further investigation was made in that respect.
On the other hand, differences between the implementations become obvious when plotting the inactivation cross sections for $\alpha$-particles with energies below 1 MeV/n. 
The relative difference between $E_\mathrm{i}$ and $E_\mathrm{f}$ gets high enough to justify the importance of considering the ion's energy loss in the sensitive volume.

We observe on Figure~\ref{subfig:sigma_comparison} that for the highest energies the inactivation cross sections obtained with the hadrontherapy and low-energy implementations are identical, as expected, since the track-segment approximation is valid over the whole cell at these energies. Moreover, inactivation cross sections decrease as the ion's energy increases because of the LET reduction. The behavior at energies lower than 1~MeV/n is more complex: 
\begin{itemize}
    \item For the hadrontherapy implementation, a saturation is observed, which is coherent with the $\frac{dn^*}{dx}$ plot (Figure~\ref{subfig:dn_de_let}). The number of lethal events per impact is $\frac{dn^*}{dx}$ $\cdot$ 1 µm, because the LET is constant with the track-segment approximation. We could expect the inactivation cross section to be equal to the geometrical one. However, this is not the case, probably because of the saturation of the NanOx lethal function.
    \item For the low-energy implementation, the variation of $n^*$ as a function of energy is taken into account. The decrease of inactivation cross sections is linked to two phenomena. On the one hand, $\frac{dn^*}{dE}$ decreases when the kinetic energy is decreasing as well. On the other hand, low-energy ions stop in the nucleus and deposit all their kinetic energy. This means that decreasing the ion's energy lead to a reduction of lethal events and therefore of the inactivation cross sections. Moreover, when the length of sensitive volume increases, the density of targets decreases (since the total number of targets is kept constant). Hence, for lower energy ions that stop in nuclei, a longer sensitive volume induces a smaller number of lethal events. In the case of ions that cross a cell nucleus without important energy loss, the number of lethal events stays constant. The negligible impact of the sensitive volume length was observed for energies above 1 MeV/n, in agreement with the study of Monini et al. (2018)\cite{monini2018}.
\end{itemize}

As mentioned earlier, the scarcity of experimental data on the radiobiological coefficient $\alpha$ at very low energies lead us to advance several hypotheses about its behavior below 0.1 MeV/n. This coefficient translates the efficiency of a particle to kill a cell, and it could be different from zero even for energies close to zero. Also, an increase of this efficiency could be observed at very low energy \cite{norarat2009very, rovituso2017nuclear}. The choice of the hypothesis on the $\alpha$ coefficient behavior under 0.1 MeV/n has no major impact in our results for $\alpha$-particles with initial kinetic energies above 1 MeV/n, irrespective of the target length (\Cref{subfig:sigma_comparison}). The $\sim$60\% differences between the chosen hypotheses, observed around 0.125 MeV/n in the cell inactivation cross sections curves correspond to only a fraction of energy deposited by the ion. However, in BNCT the energies of helium and lithium ions are between 0.12 and 0.45 MeV/n. In that context, careful attention must be paid to the interpretation of the NanOx model predictions obtained with the present implementation. Indeed, it is likely that further improvements are needed to adapt the NanOx model to BNCT applications. In all cases, validation against available experimental data will be required.

\subsection{NanOx within the state-of-the-art biophysical models}
The idea behind the NanOx developments is to be as close as possible to the physical reality. The stochastic nanometric aspect of NanOx, as well as the consideration of free radicals is designed with this in mind, with the goal of predicting cell survivals and RBE in a detailed and novel way. \\
The NanOx implementation described in this paper is built to be coherent between energy scales: the same lethal function is used for hadrontherapy applications and for low-energy ion irradiations. Our purpose with NanOx is to provide a consistent long-term mechanistic approach for describing the biological effects induced by ionizing radiation. This is in contrast to more pragmatic approaches, such as the one found in the MIRD recommendations \cite{sgouros2010mird}. Moreover, instead of assuming cell death when 2 $\alpha$-particles cross the nucleus, NanOx can add more precision and granularity in cell survival calculations. In recent years, other biophysical models offering micrometric resolution in the dosimetry of high-LET radiations have been reported in the literature. Some notable examples include the stochastic microdosimetric kinetic (SMK) model \cite{sato2012,sato2018} and the integrated microdosimetric kinetic (IMK) model \cite{matsuya2018,matsuya2020}. The SMK model takes into account the stochastic nature of specific energy in the whole cell nucleus, as well as in each microscopic site within the nucleus, the so-called domains. The IMK model was developed to assess the DNA repair during irradiation as well as non-targeted effects after low-dose acute irradiation. Both models are currently able to include other aspects of irradiation such as dose rate and oxygen effects in their calculations. The quantitative comparison of NanOx with other biophysical models is outside the scope of the present manuscript, and will be explored in another paper.

\subsection{Extending the model to other biological targets}
In this study, only one sensitive volume was considered, i.e. the cell nucleus. In the literature the role of irradiated extranuclear targets on cell's fate has been observed and investigated \cite{pouget2022target, pouget2021revisiting}. $\alpha$-particles can cross several cells before stopping. Lethal nucleus damage can then be observed as the main contribution to cell death \cite{pugliese1997inactivation}. However in BNCT, depending on the intracellular distribution of the boron compound and of the emitted low-energy ions, it could not be the case. With a range of just a few µm, an important part of the ions emitted in BNCT might not traverse the nucleus. This part depends on the solid angle and would be larger if the boron compound did not enter the cell's cytoplasm or nucleus. Still, even without depositing energy in the nucleus, the ions may damage the cytoplasm and the cell membrane, potentially activating pathways leading to cell death or mutation. The modeling of extranuclear damage within the NanOx framework is currently in progress.


\newpage
\section{Conclusions}
\label{sec:conclusion}
This paper presents the extension of the NanOx model to consider irradiations with low-energy ions, typical of innovative radiotherapies such as TAT and BNCT. Some approximations were introduced to facilitate the mathematical formulation of the problem and to reduce computation time. Comparisons were performed, in terms of inactivation cross sections, against the previously reported hadrontherapy implementation of NanOx. The agreement between the latter and the low-energy implementation of NanOx is consistent in the energy range in which they are expected to overlap (e.g. for energies above 1 MeV/n), and remarkable for relatively high energies (e.g. 50 MeV/n). Further work is still needed to validate experimentally the low-energy implementation of the NanOx model and to extend it to multiple sensitive volumes.


\section*{Acknowledgments}
This work was performed in the framework of the LabEx PRIMES (ANR-11-LABX-0063) of the Université de Lyon, within the program ``Investissements d’Avenir'' (ANR-11-IDEX-0007) operated by the French National Research Agency (ANR). We acknowledge the financial support of the French National Institute of Health and Medical Research (Inserm), through the grant ``Apports à l’oncologie de la physique, de la chimie et des sciences de l’ingénieur'', no. 20CP176-00.

\section*{Conflict of Interest Statement}
All authors declare that they have no conflicts of interest to disclose.

\appendix
\section*{Appendices}
\label{sec:appendices}
\addcontentsline{toc}{section}{\numberline{}Appendices}

\setcounter{figure}{0}  
\renewcommand{\thesubsection}{\Alph{subsection}}
\renewcommand\thefigure{\thesubsection.\arabic{figure}}  

\subsection{Energy variation in the sensitive volume}
\label{sec:appendix_eief}
\begin{figure}[H]
\centering
\vspace{5mm}
\includegraphics[height=8cm]{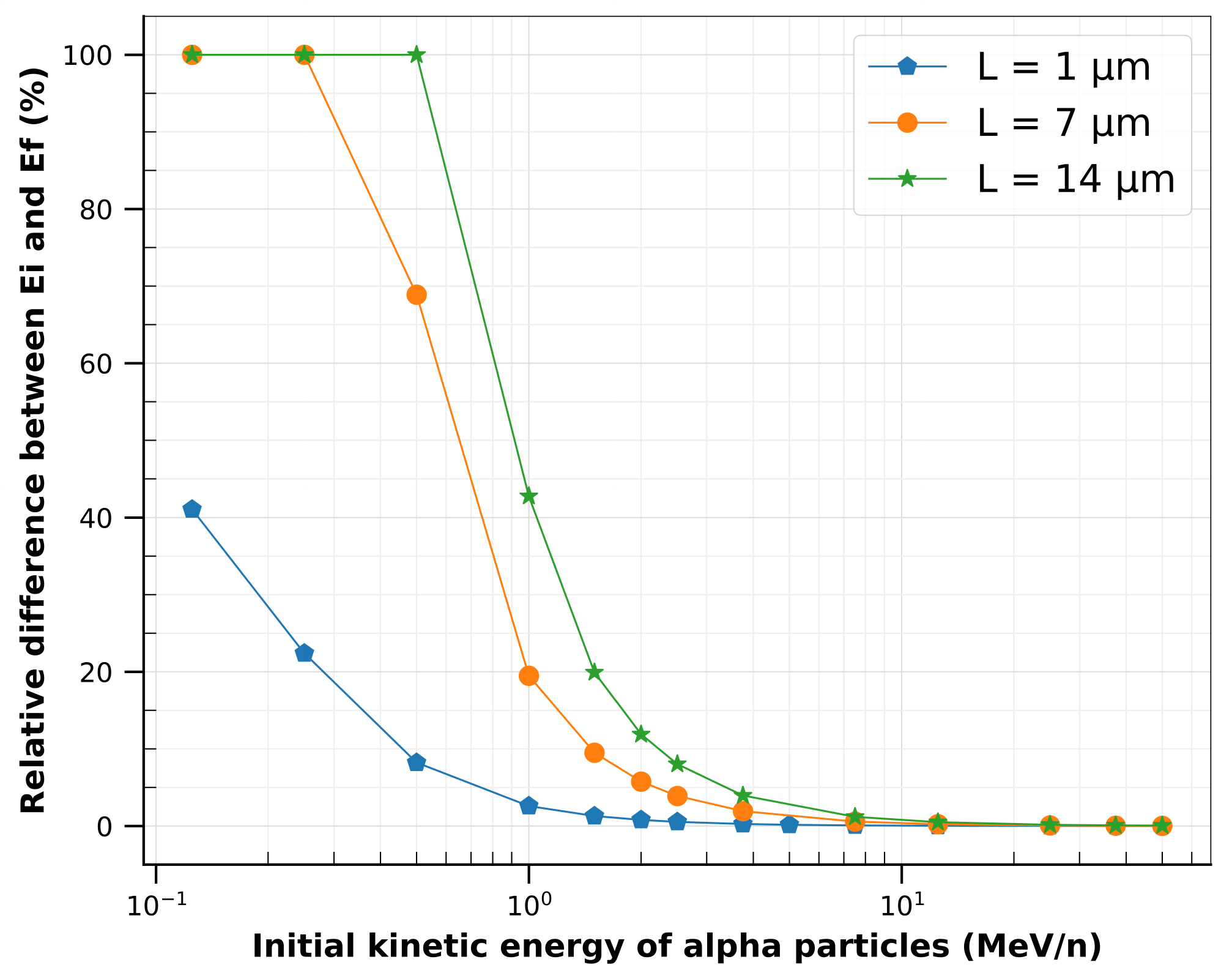}
\caption{Mean relative difference $\frac{E_i - E_f}{E_i}$ as a function of the initial kinetic energy of $\alpha$-particles. The simulation conditions were the same as in \Cref{subsec:inactivationcrosssection_results}.}
\label{fig:appendix_ei_ef_lost}
\end{figure}

Computing the energy lost by ions when traversing the sensitive volume is interesting for evaluating the effect of geometry (e.g. sensitive volume length) in our assumptions and to determine the domain of application of the low-energy NanOx implementation. To this end, \Cref{fig:appendix_ei_ef_lost} presents the relative difference between the energies $E_{\mathrm{i}}$ and $E_{\mathrm{f}}$ as a function of the initial kinetic energy of $\alpha$-particles and for different sensitive volume lengths.

\subsection{Link between $\frac{\Xk{\dd n^*}{}{}}{\Xk{\dd E}{}{}}$ and $\alpha$ at intermediate energies}
\label{sec:appendix_dnde_alpha}

Let us consider a cell irradiated by a parallel and homogeneous beam of monoenergetic ions. For ions of intermediate energies, \ie energies low enough to assume that the tracks are narrow, but high enough to assume that ions follow straight line trajectories with constant velocity when traversing the sensitive volume, the integration of \Cref{eq:tNtkn_integral} is easy since the term $^{\tN, \tk}\left(\frac{ \Xk{\dd n^*}{}{}}{\dd E}\right)$ can be approximated as constant. For instance, the order of magnitude of this energy range for an $\alpha$-particle crossing 1 µm of water would be between 1 and 10 MeV/n (\Cref{fig:appendix_ei_ef_lost}).

For narrow tracks, the inactivation cross section $\sigma$ becomes:
\begin{equation}
\Xk{\sigma}{\tk}{} = \sigma_{\mathrm{s}}(1 - \Xk{\langle S_1 \rangle}{\tk}{})\ ,
\label{eq:sigmatk}    
\end{equation}

where $\sigma_{\mathrm{s}}$ is the geometrical cross section of the sensitive volume (i.e. the cell nucleus). The radiobiological $\alpha$ coefficient is given by:
\begin{equation}
\Xk{\alpha}{\tk}{} = \frac{\Xk{\sigma}{\tk}{}}{a \cdot \Xk{\mathrm{LET}}{\tk}{}} \ ,
\label{eq:alphatk}    
\end{equation}

where the factor $a =$ \SI{0.1602}{\Gy\nkeV\cbum} is used to convert units; and the linear energy transfer in \si{\keVum} is equal to $\Xk{\mathrm{LET}}{\tk}{} =~ ^{\tk}\left(\frac{\Xk{\dd E}{}{}}{\dd x}\right)\cdot \mathrm{CPE}$, with $\mathrm{CPE}$ a factor related to the condition of charged-particle equilibrium. We currently set $\mathrm{CPE} = 1$.

If one approximates the cell nucleus with a cylindrical geometry and neglects the fluctuations during the transversal of the cell nucleus, we have from \Cref{eq:tNtkn_integral}:
\begin{equation} 
\Xk{n^*}{\tN, \tk}{} =  
\int^{\Xk{E_{\mathrm{i}}}{\tk}{}}_{\Xk{E_{\mathrm{f}}}{\tk}{}} {\prescript{\tN,\tk}{}{\left(\frac{\Xk{\dd n^*}{}{}}{\Xk{\dd E}{}{}}\right)} \dd E} \approx
\prescript{\tN,\tk}{}{\left(\frac{\Xk{\dd n^*}{}{}}{\Xk{\dd E}{}{}}\right)}
\left(\Xk{E_{\mathrm{i}}}{\tk}{} - \Xk{E_{\mathrm{f}}}{\tk}{}\right)
\ , 
\label{eq:tNtkn_approx} 
\end{equation}

with:
\begin{equation} 
\left(\Xk{E_{\mathrm{i}}}{\tk}{} - \Xk{E_{\mathrm{f}}}{\tk}{}\right) 
\approx
\prescript{\tk}{}{\left(\frac{\Xk{\dd E}{}{}}{\dd x}\right)}
\cdot L_{\mathrm{s}}\ ,
\label{eq:EiEf_approx} 
\end{equation} 

where $L_{\mathrm{s}}$ is the length of the cylinder representing the cell nucleus. 

Similarly, for global events, the integral in \Cref{eq:Yintegral} is simplified as follows:
\begin{equation} 
\frac{\eta}{m_{\mathrm{s}}}\int^{\Xk{E_{\mathrm{i}}}{\tk}{}}_{\Xk{E_{\mathrm{f}}}{\tk}{}} {\frac{\Xk{G(E)}{\tk}{}}{\Gr}~\dd E} =
\frac{\eta}{m_{\mathrm{s}}}\cdot\frac{\Xk{G(E)}{\tk}{}}{\Gr}\cdot
\prescript{\tk}{}{\left(\frac{\Xk{\dd E}{}{}}{\dd x}\right)}
\cdot L_{\mathrm{s}}
\ .
\label{eq:G_int_approx} 
\end{equation} 

The number of lethal events for one impact is then written as:
\begin{equation}
\begin{split}
\Xk{n}{\tN, \tk}{1} &=
\prescript{\tN,\tk}{}{\left(\frac{\Xk{\dd n^*}{}{}}{\Xk{\dd E}{}{}}\right)}
\cdot 
\prescript{\tk}{}{\left(\frac{\Xk{\dd E}{}{}}{\dd x}\right)}
\cdot L_{\mathrm{s}}
+ \aG\cdot\frac{\eta}{m_{\mathrm{s}}}\cdot\frac{\Xk{G(E)}{\tk}{}}{\Gr}\cdot 
\prescript{\tk}{}{\left(\frac{\Xk{\dd E}{}{}}{\dd x}\right)}
\cdot L_{\mathrm{s}}\\
&+ \bG\cdot\left[\frac{\eta}{m_{\mathrm{s}}}\cdot\frac{\Xk{G(E)}{\tk}{}}{\Gr}\cdot 
\prescript{\tk}{}{\left(\frac{\Xk{\dd E}{}{}}{\dd x}\right)}
\cdot L_{\mathrm{s}}\right]^2
\ . 
\end{split}
\label{eq:n1tktN} 
\end{equation} 

The survival for one impact is then given by:
\begin{equation}
\Xk{\langle S_1 \rangle}{\tN, \tk}{} =~ \mathrm{exp}\left(-\Xk{n}{\tN, \tk}{1}\right) \ .
\label{eq:S1n1} 
\end{equation}

From \Cref{eq:sigmatk,eq:alphatk,eq:S1n1} we can derive the following expressions for $\Xk{\sigma}{\tk}{}$ and $\Xk{\alpha}{\tk}{}$:
\begin{equation} 
\Xk{\sigma}{\tk}{} = \sigma_{\mathrm{s}}\left[1 - \mathrm{exp}\left(-\Xk{n}{\tN, \tk}{1}\right) \right]
\ , 
\label{eq:sigma_n1} 
\end{equation}
\begin{equation} 
\Xk{\alpha}{\tk}{} = \frac{\sigma_{\mathrm{s}}\left[1 - \mathrm{exp}\left(-\Xk{n}{\tN, \tk}{1}\right) \right]}{a\cdot\Xk{\mathrm{LET}}{\tk}{}}
\ . 
\label{eq:alpha_n1} 
\end{equation} 

It can be seen that \Cref{eq:alpha_n1} provides a simple link
between the density of lethal events per keV and the radiobiological coefficient $\alpha$ usually tabulated for hadrontherapy applications.

\section*{References}
\addcontentsline{toc}{section}{\numberline{}References}
\vspace*{-10mm}





\bibliography{./references}      



\bibliographystyle{./medphy.bst}    




\end{document}